\documentclass[conference]{IEEEtran}
\IEEEoverridecommandlockouts
\usepackage{cite}
\usepackage{amsmath,amssymb,amsfonts}
\usepackage{acronym}
\usepackage{algorithmic}
\usepackage{float}
\usepackage{graphicx}
\usepackage{textcomp}
\usepackage{xcolor}
\usepackage{url}
\def\BibTeX{{\rm B\kern-.05em{\sc i\kern-.025em b}\kern-.08em
    T\kern-.1667em\lower.7ex\hbox{E}\kern-.125emX}}
\usepackage[numbers]{natbib}
\usepackage{tikz}
\usetikzlibrary{decorations.text}
\usetikzlibrary{shapes}
\usepackage{pgfplots}
\usepackage{pgfplotstable}
\usepgfplotslibrary{groupplots}
\usepgfplotslibrary{fillbetween}
\usepackage{subfig}
\pgfplotsset{compat=1.16}
\usepackage{mathabx}
\usepackage{flushend}
\begin{document}

\title{HiPARS: Highly-Parallel Atom Rearrangement Sequencer
\thanks{This work was funded by the German Federal Ministry of Education and Research (BMBF) under the funding program \textit{Quantum Technologies - From Basic Research to Market} under contract number 13N16087, as well as from the Munich Quantum Valley~(MQV), which is supported by the Bavarian State Government with funds from the Hightech Agenda Bayern.}
}

\author{\IEEEauthorblockN{Jonas Winklmann}
\IEEEauthorblockA{
\textit{Technical University of Munich}\\
\textit{TUM School of Computation, Information and Technology}\\
\textit{Chair of Computer Architecture and Parallel Systems}\\
Munich, Germany\\
jonas.winklmann@tum.de}
\and
\IEEEauthorblockN{Martin Schulz}
\IEEEauthorblockA{\textit{Technical University of Munich}\\
\textit{TUM School of Computation, Information and Technology}\\
\textit{Chair of Computer Architecture and Parallel Systems}\\
Munich, Germany\\
schulzm@in.tum.de}
}

\acrodef{EMCCD}[EMCCD]{electron-multiplying charge-coupled device}
\acrodef{CMOS}[CMOS]{complementary metal-oxide-semiconductor}
\acrodef{qCMOS}[qCMOS]{quantitative \ac{CMOS}}
\acrodef{MPQ}[MPQ]{Max Planck Institute for Quantum Optics}
\acrodef{sCIC}[sCIC]{serial clock-induced charge}
\acrodef{CIC}[CIC]{clock-induced charge}
\acrodef{SNR}[SNR]{signal-to-noise ratio}
\acrodef{EM}[EM]{electron-multiplying}
\acrodef{MTF}[MTF]{modulation transfer function}
\acrodef{PSF}[PSF]{point-spread function}
\acrodef{OTF}[OTF]{optical transfer function}
\acrodef{CDF}[CDF]{cumulative distribution function}
\acrodef{PDF}[PDF]{probability density function}
\acrodef{GCC}[GCC]{GNU Compiler Collection}
\acrodef{ROI}[ROI]{region of interest}
\acrodef{ADC}[ADC]{analog-to-digital converter}
\acrodef{MQV}[MQV]{Munich Quantum Valley}
\acrodef{PSF}[PSF]{point spread function}
\acrodef{RL}[RL]{Richardson-Lucy}
\acrodef{AOD}[AOD]{acousto-optic deflector}
\acrodef{SLM}[SLM]{spatial light modulator}
\acrodef{LSAP}[LSAP]{Linear Sum Assignment Problem}
\acrodef{HPFA}[HPFA]{heuristic path-finding algorithm}
\acrodef{HCA}[HCA]{heuristic cluster algorithm}
\acrodef{RF}[RF]{radio frequency}
\acrodef{PSCA}[PSCA]{parallel sort-and-compression algorithm}
\acrodef{ASA}[ASA]{A* search algorithm}

\bibliographystyle{IEEEtranN}

\IEEEpubid{\begin{minipage}[t]{\textwidth}\ \\[10pt]
\centering
979-8-3315-5736-2/25/\$31.00 \copyright 2025 IEEE. Personal use of this material is permitted. Permission from IEEE must be obtained for all other uses, in any current or future media, including reprinting/republishing this material for advertising or promotional purposes, creating new collective works, for resale or redistribution to servers or lists, or reuse of any copyrighted component of this work in other works. DOI 10.1109/QCE65121.2025.00065
\end{minipage}}

\maketitle

\begin{abstract}
Neutral atom quantum computing's great scaling potential has resulted in it emerging as a popular modality in recent years. For state preparation, atoms are loaded stochastically and have to be detected and rearranged at runtime to create a predetermined initial configuration for circuit execution. Such rearrangement schemes either suffer from low parallelizability for \ac{AOD}-based approaches or are comparatively slow in case of \acp{SLM}. In our work, we introduce an algorithm that can improve the parallelizability of the former. Since the transfer of atoms from static \ac{SLM} traps to \ac{AOD}-generated movable traps is detrimental both in terms of atom loss rates and execution time, our approach is based on highly-parallel composite moves where many atoms are picked up simultaneously and maneuvered into target positions that may be comparatively distant. We see that our algorithm outperforms its alternatives for near-term devices with up to around 1000 qubits and has the potential to scale up to several thousand with further optimizations.
\end{abstract}


\section{Introduction}
\begin{figure*}[!t]
\centering
  \subfloat[]{\includegraphics[width=0.16\linewidth]{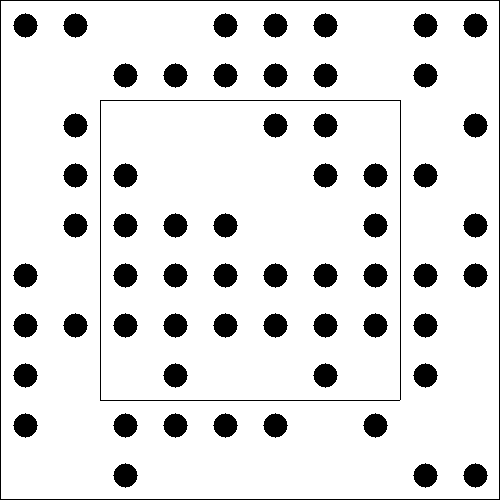}\label{figures:startingPos}}
  \hfill
  \subfloat[]{\includegraphics[width=0.16\linewidth]{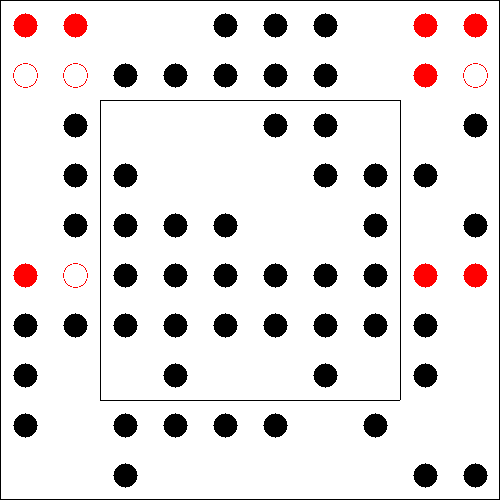}}
  \hfill
  \subfloat[]{\includegraphics[width=0.16\linewidth]{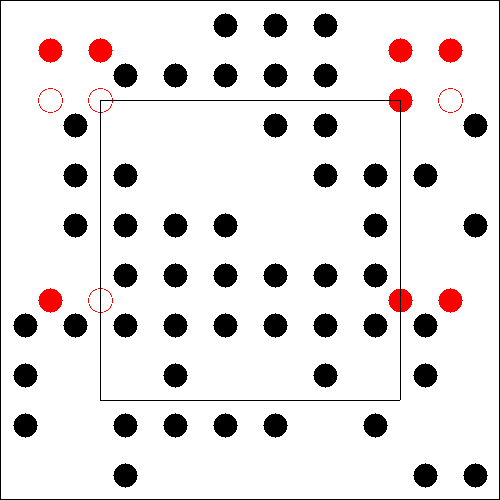}}
  \hfill
  \subfloat[]{\includegraphics[width=0.16\linewidth]{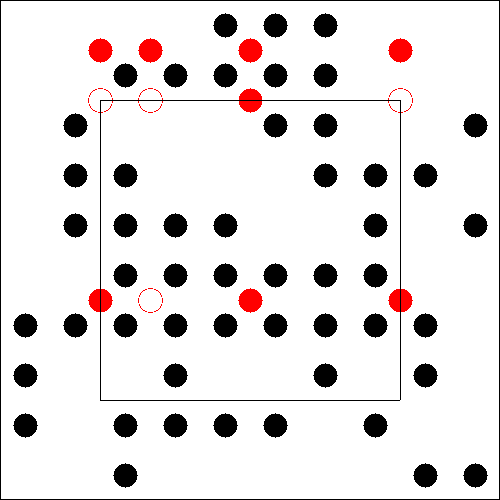}}
  \hfill
  \subfloat[]{\includegraphics[width=0.16\linewidth]{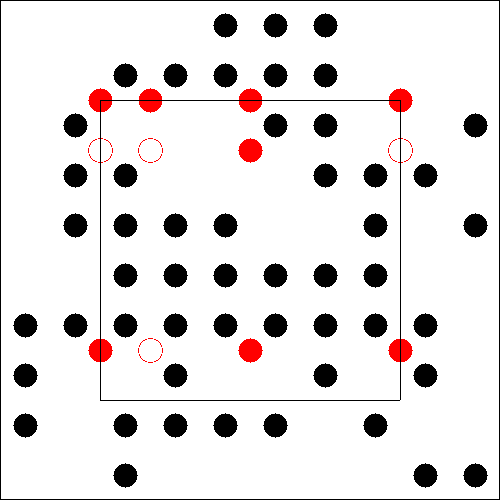}}
  \hfill
  \subfloat[]{\includegraphics[width=0.16\linewidth]{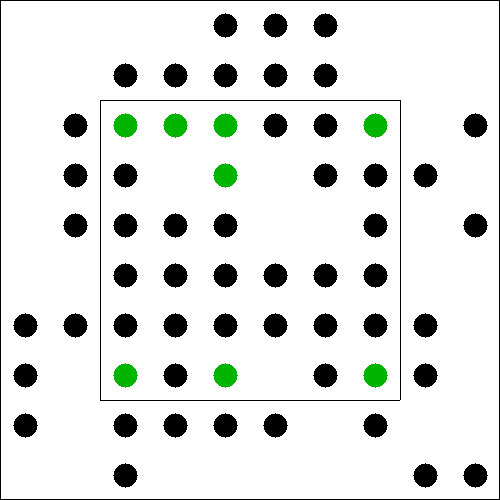}}
\caption{Composite move consisting of 4 steps during rearrangement from 10x10 whole array to centered 6x6 target area (surrounded by black border). Black circles represent atoms, filled red circles represent atoms that are used during move, Empty red circle shows positions of empty moving traps. (a) Starting position. (b) Selection of atoms. (c) - (e) State after steps 1-3. (f) Final positions after last step.}\label{figures:moveSequence}
\end{figure*}
Neutral atom quantum computing has recently emerged as a popular modality, especially due to its high scaling potential, and several companies and research groups are on the verge of operating around the 1000-qubit range \cite{Pichard:1,Pause:1}. These setups usually require an atom array around twice the target size, which is stochastically loaded so that approximately 50\% of sites contain an atom. These are then detected through fluorescence imaging~\cite{Morgado:1} and rearranged to produce a predefined target area, which is often a sub-area where every site is occupied. For atom counts in the thousands, there is not yet a publicly-known best method for rearranging the stochastically-loaded atoms into a deterministic pattern. \Acf{AOD}-based rearrangement schemes mostly still suffer from low parallelizability, while newly emerging \acf{SLM}-based schemes solve this problem at the cost of a comparatively slow scaling per measure of distance of the longest move. These shortcomings lead to the current situation where the compute cycle time of many setups is dominated by initialization steps, especially the sorting \cite{Quera:1,Gyger:1}.

A lot of work has been done to optimize the underlying physics problem of transporting the atoms \cite{Lam:1,Murphy:1,Torrontegui:1}. However, there is also the software problem of finding a fast movement sequence that is gaining more attention. While some efforts have been undertaken to parallelize \ac{AOD}-based setups, all current solutions solve the complex problem of finding moves by employing very simple heuristics, like compactifying all atoms in a row in one move. This leaves a lot of potential speedup that is still to be gained and we believe that a more complex approach that generates highly-parallel composite moves can realize much of it.


Our solution to the problem of finding a rearrangement sequence is to develop a heuristic greedy algorithm that executes the best possible move it can find at each step. To do so, there are several methods suggesting different types of moves that may be suited for different phases of the sorting process. These methods themselves use a configurable cost function to determine the time it takes for the move to be executed. This same function is also used to choose the best overall move to be executed. In order to minimize time spent on atom sorting, we choose the rearrangement step that maximizes the quotient of the number of target sites that executing the move would fill and the time required to execute it. The moves that may be suggested at every step fulfill different demands. There are methods for suggesting complex parallel moves as well as ones that provide faster direct moves, should the situation require them.
\IEEEpubidadjcol

In order to allow for an easier understanding of our algorithm, we will follow this section by describing the underlying hardware and how the rearrangement procedure typically works. After that, we will describe the intricacies of our novel approach and the results it can achieve, followed by the best existing alternatives. Having done so, we can compare our algorithm to these alternatives in section~\ref{section:eval} in order to see whether we can offer meaningful advantages. Having done so, we conclude by summarizing the performance of our algorithm and the situations in which it excels.

In summary, our contributions include:
\begin{itemize}
    \item Investigating movement heuristics that are flexible enough to be able to suggest both long and complex moves as well as fast ones.
    \item Introducing a configurable cost function that penalizes moves based on actual time spent on execution in order to select the best possible choice of move at any step during rearrangement.
    \item Finely tuning parameters to balance algorithm run time and sophistication of movement procedure.
\end{itemize}

We found that our approach can save more than 50\% of the move-execution time of the currently best methods. While run time is currently a limitation, our algorithm can, even if not further optimized, provide meaningful speedups to the rearrangement procedure for all suitable setups with qubits numbers up to around 1000.
\section{Background}
We generally consider the case where there exists a setup where there is a rectangular grid of regularly-spaced stationary traps, as well as additional movable ones that are generated using two \acp{AOD} oriented perpendicular to each other. We call them the horizontal and vertical \ac{AOD} and we can feed each one a certain number of independent \ac{RF} tones. We will refer to the limit of the numbers of possible tones by $n_h$ and $n_v$ respectively. By altering the frequencies, we can steer the movable traps and, thereby, transport atoms around the working area.

The first step in initializing a neutral atom quantum computer for operation is to load atoms into the stationary traps. This process is intrinsically stochastic in the sense that we can only know the approximate probability of each site containing an atom. Since we want to start circuit execution from a predetermined arrangement of atoms, we need to move some from positions where they are not required to unoccupied sites which need to be filled. Often, this desired predetermined configuration is defined as a subarray where every site contains an atom. We will refer to the total number of atoms in the target area as $n_t$ and we will generally use square arrays, although this is not a requirement.

As mentioned before, some groups and companies have already or are about to deal with system with $n_t\geq 1000$. For cases like these, the total time spent on rearrangement including calculations and move execution can exceed one second. While neutral atom quantum computing is not known for being fast, but rather despite its rather slow compute cycle, improving this is quite important.

To facilitate understanding of how rearrangement works, Fig.~\ref{figures:moveSequence} shows an exemplary move based on twelve moving traps of which eight are occupied. Starting from the randomly occupied array in Fig.~\ref{figures:startingPos}, three rows and four columns are chosen and all atoms that have both their row and column selected are moved simultaneously. Going through the illustrations, we can see how we start by moving the traps half a step diagonally in order to allow for them to be moved within the gaps between rows and columns in the following steps. Lastly, they are moved by another half-step to reach their target locations. Within the centered target area, which we illustrated by surrounding it with a black border, there are initially 13 empty sites of which we can fill eight within a single move. Using a sequential sorting technique would require at least 13 moves while we require three in this case.
\section{Rearrangement Algorithm}
In the following, we will explain the intricacies and performance of our sorting algorithm. It is mostly implemented using C++ to ensure fast run times and is fitted with a Python wrapper to allow for easy integration into existing software. It is publicly available via GitHub \cite{GitHub:1}.

To begin with, it is to be noted that this algorithm is, at the moment, exclusively capable of rearranging into a fully occupied configuration, i.e, where no site within the target area is empty. However, it should be very well suited for this case, since the moves it generates can transport a lot more atoms simultaneously than other known approaches. To find a movement sequence for a given case, it accepts the initial occupancy matrix of the whole array as well as the location and size of the target subarray and returns a list of rearrangement steps, which we call moves, if rearrangement is possible.

The constraint that limits the parallelizability of moves most and that makes finding moves quite complicated is the fact that for a set of vertical and horizontal tones $f_h=\{f_{h,1},...,f_{h,i}\}$ and $f_v=\{f_{v,1},...,f_{v,j}\}$ with $i<n_h$ and $j<n_v$, there will be traps at every possible set of coordinates in the Cartesian product $f_h\bigtimes f_v$. We can see this in Fig.~\ref{figures:moveSequence}, where we only require the filled red circles. However, there are also traps at the locations of the empty red circles. For every generated moveable trap, we have to take care that start and end position cannot both be filled and that we do not cross over any occupied sites on the way. We also allow the user to configure that the trap at the final position has to be empty, even if the start trap was already unoccupied.

Since we can freely choose which rows and columns to address up to the maximum number of \ac{RF} tones, there are $(\sum_{i=0}^{n_h}{\sqrt{n_t}\choose i}^2))\cdot(\sum_{i=0}^{n_v}{\sqrt{n_t}\choose i}^2))$ possible moves to consider at any point, given an atom array of size $\sqrt{n_t}\bigtimes\sqrt{n_t}$. For $\sqrt{n_t}=50$ and $n_h=n_v=16$, which are reasonable numbers for our case, there are already approximately $9.2\cdot 10^{50}$ possibilities. The vast majority of these moves will not be allowed, but finding moves that are allowed and useful is not a trivial problem.
\subsection{Configuration}
Between different systems, spacings between atoms as well as characteristics of lasers and other hardware may differ, which imposes different constraints on our rearrangement sequence. The result of this is that different moves may be allowed and we may want to tune the balance between calculation time and rearrangement execution time. To allow for careful tuning to a system's characteristics, we provide several configuration options:
\begin{itemize}
    \item Maximum number of horizontal and vertical \ac{AOD} frequencies ($n_h$ and $n_v$).
    \item Maximum number of movable traps. Since laser power may be the limiting factor, we want to be able to constraint the total number of traps that are generated at any given step, i.e, this is a limit to the effective $n_h\cdot n_v$.
    \item Is movement between rows and columns possible? While this can be configured, disallowing either of these two movement options severely limits the parallelization capabilities of our approach.
    \item Complexity Tuning: Maximum number of combinations of rows and columns to consider for parallel moves
    \item Time-demand function: A function describing the time it takes to physically execute a move. It serves as a cost function for selecting moves. To simplify, a constant offset, a constant per-substep offset, a factor scaling linearly with distance, and a factor scaling with the square root of the distance can be configured.
    \item May atoms be moved several times? As movement may heat atoms, we want to be able to disallow multiple moves per atom.
    \item Is it allowed to move an empty trap onto an occupied final position?
\end{itemize}
Additionally, the algorithm accepts a boolean array of the initial occupancy as well as the location and size of the target area. After finishing calculations, it will return a list of generated moves if sorting was successful. Otherwise, it will return nothing.
\subsection{Time-demand function}\label{section:tdf}
Different rearrangement schemes may be based on different balances of move count and distances. Finding a function to simulate the move-execution time will never be truly fair to all of them, since different time demands of a system's transport capabilities may favor some. However, if we want to meaningfully compare them, we need to establish a representative function that describes the time it takes to execute our generated sequence. 

A move starts by simultaneously ramping up the amplitudes of a set of \ac{RF} tones for the vertical and horizontal \ac{AOD} in order to transfer atoms from the stationary traps to the movable tweezers. The frequencies are then modified to execute a certain number of substeps until the target positions are reached. At that point, the amplitudes are ramped down again to transfer the atoms back to their stationary traps. For each substep, one could argue about the way of interpolating between start and end position. However, we will see in Section~\ref{section:relatedWork} that some other sources only state total distance of their schemes. If we want to meaningfully compare to them, then our function describing a move's time demand ought to only reference total distance linearly. This does not constitute a problem since additional time spent on accelerating and decelerating can be attributed to the constant loading time. Exemplary maximum transfer speeds are 54$\frac{\mu m}{ms}$ \cite{Gyger:1}, 130$\frac{\mu m}{ms}$ \cite{Tian:1}, and 550$\frac{\mu m}{ms}$ \cite{Bluvstein:1}. Our function, therefore, contains a constant offset for transferring the atoms from and to the stationary grid as well as a component that scales linearly with the distance of the move, representing the actual shuttling of the atoms. Different sources for the constant portion state values at 60-400$\mu s$ for start and end of the move each \cite{Tian:1,Bluvstein:2,Gyger:1}. 

As our time-demand function mainly aims at being representative, we looked at values used in other sources. Since \citeauthor{Tian:1} do explicitly mention both their values and both are realistic, we will also use their function \cite{Tian:1}. For a move with $c$ substeps, $j$ total \ac{RF} tones, and $d_{n,m}$ denoting the distance traversed by tone $m$ during substep $n$, we define the total distance $d$ as the sum of the longest distances of each submove.
\begin{equation}
    d=\sum_{i=0}^c{max(d_{i,0},...,d_{i,j})}
\end{equation} 
Using that distance we state the time-demand function for executing a move:
\begin{equation}
    t_m(d)=120\mu s+\frac{d}{0.13\frac{\mu m}{\mu s}}
\end{equation}
\subsection{Solution}
Our algorithm is a heuristic greedy algorithm that, at any given step during the rearrangement procedure, will investigate different types of moves and execute the best one. Considering the vast number of possible moves at any given step, we deem it impractical to consider them all. Our method is a fresh approach compared to previous approaches both in the way it finds moves and how it evaluates them.

It uses a configurable fitness function to evaluate different moves and selects the best one for execution. This function is currently defined by dividing the number of target positions a move fills by the time taken to execute it. It may also be useful to incorporate fidelity into the function or to weight target positions in order to rewards moves that fill difficult-to-reach locations.

During the generation of potential moves, the algorithm produces different types of moves that adhere to different heuristics. The generation of each already takes this fitness function into consideration to produce an effective rearrangement step. The currently implemented types of moves, ordered from relatively fast to more complicated, are:
\begin{itemize}
    \item Direct compactification move: Compactifying the tones in a given row or column. This move is comparatively fast, especially if the move cost is heavily dependent on the square root of the distance, but only fills few target positions.
    \item Moving a row or column laterally into the target area, e.g, moving each tone in a row half a step horizontally, then vertically into the target area, and another horizontal half-step into position.
    \item Moving a row or column lengthwise into target area, e.g, moving a row vertically by half a step, then horizontally into the target area, and another vertical half-step into position.
    \item Complex multi row/column move. Select several source rows and columns for which there exists a target set of rows and columns such that there is no case where both source and target location are occupied.
\end{itemize}
Naturally, the last-mentioned type of move is by far the most complicated and can take quite long to calculate. However, it is also the most important one as it is the only one that, at this stage, is capable of producing a move that utilizes several rows and columns simultaneously. Since run time may be an issue in some situations, we introduce a configuration option to limit the number of states that may be investigated here.

Each of the functions suggesting these different types of moves is called after every move to return suggestions for the next one. These suggestions are then evaluated again to produce the move that maximizes the number of filled target positions per measure of time that would need to be spent to execute it. After selecting the best move, further steps are undertaken to optimize it. Ideally, one would conduct this procedure with every considered move. However, timing constraints prohibit this. There are, at the moment, two optimization passes.

The first iterates over all possible source rows and columns that are currently not used and checks whether there is a target row or column that would be compatible with the move. If that is the case, the one that offers the best improvement according to our filled-positions-per-time function is selected and added to the move. This procedure is repeated until no viable pair of source and target rows or columns can be found that improves the move. While this optimization can improve many rearrangement steps greatly, it is practically useless towards the end of the sorting procedure when many remaining unoccupied target sites have neither a coinciding row nor column. Let us consider the example where we have to fill sites at coordinates (0/0) and (1/1) and we have a move that fills the former, then it cannot be possible to find a single additional row or column that can be added to also fill the latter, so the optimization would return immediately. To remedy this shortcoming, the second optimization step finds such independent sites and tries to add an additional row and column frequency to fill them simultaneously. As stated before, this is especially important towards the end of the sorting procedure which can otherwise become almost completely sequential. These optimization steps are a vital part in providing highly parallel moves, since only one of the move types uses multiple rows and columns without optimization.
\section{Results}
\pgfplotsset{height=0.38\textwidth,width=\axisdefaultwidth,compat=1.9}
\begin{figure}
\centering
\begin{tikzpicture}
\begin{axis}[
    xlabel={Target Atom Count},
    ylabel=Number of moves,
    legend pos=north west,ymin=0,xmin=0,xmax=3700,
    legend entries={Default,$t_{m,2}$,$t_{m,2}\text{;}\textstyle n_h=n_v=8$,$t_{m,2}\text{;}k=16$,$t_{m,2}\text{;}f=0.75$,$t_{m,2}\text{;}r=2$},legend style={font=\footnotesize}
    ]
  \addplot[blue,mark=,error bars/.cd, y dir=both, y explicit] table [x=Qubit,y=AMC,y error=MCStd] {data/tD1_5-0_5-dist.dat};
  \addplot[black,dashed,mark=,error bars/.cd, y dir=both, y explicit] table [x=Qubit,y=AMC,y error=MCStd] {data/default.dat};
  \addplot[black,mark=,error bars/.cd, y dir=both, y explicit] table [x=Qubit,y=AMC,y error=MCStd] {data/tD1_5-0_5-8x8.dat};
  \addplot[brown,mark=,error bars/.cd, y dir=both, y explicit] table [x=Qubit,y=AMC,y error=MCStd] {data/tD1_5-0_5-k16.dat};
  \addplot[red,mark=,error bars/.cd, y dir=both, y explicit] table [x=Qubit,y=AMC,y error=MCStd] {data/tD1_5-0_75.dat};
  \addplot[green,mark=,error bars/.cd, y dir=both, y explicit] table [x=Qubit,y=AMC,y error=MCStd] {data/tD2-0_5.dat};
\end{axis}
\end{tikzpicture}
\caption{Average move counts that our rearrangement algorithm generates given a target atom count, a time-demand function, an initial filling ratio, the ratio $r$ between total and target size, and limiting factors to the parallelism. Error bars denote $\pm$ one standard deviation.\label{figures:moveCounts}}
\end{figure}
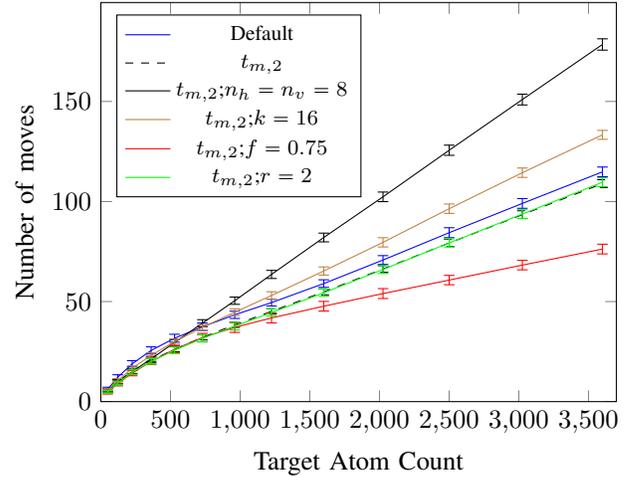

\pgfplotsset{height=0.33\textwidth,width=0.45\linewidth,compat=1.9}
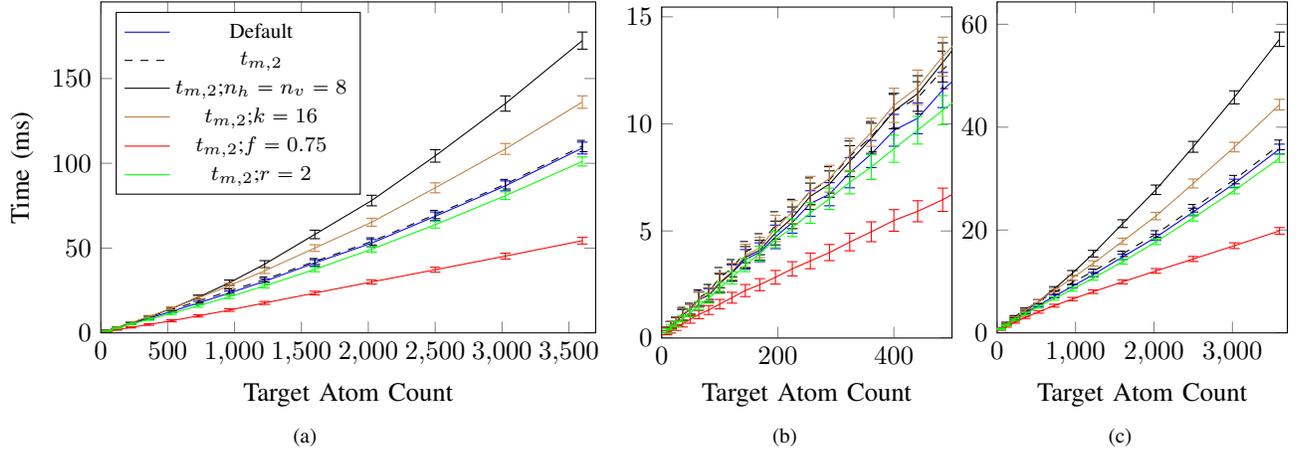
\begin{figure*}
\centering
\subfloat[]{\label{figures:executionTime13}\begin{tikzpicture}
\begin{axis}[
    xlabel={Target Atom Count},
    ylabel=Time (ms),
    legend pos=north west,ymin=0,xmin=0,xmax=3700,
    legend entries={Default,$t_{m,2}$,$t_{m,2}\text{;}\textstyle n_h=n_v=8$,$t_{m,2}\text{;}k=16$,$t_{m,2}\text{;}f=0.75$,$t_{m,2}\text{;}r=2$},legend style={font=\footnotesize}
    ]
  \addplot[blue,mark=,error bars/.cd, y dir=both, y explicit] table [x=Qubit,y expr=\thisrow{AMT13} / 1000,y error expr=\thisrow{MTStd13} / 1000] {data/default.dat};
  \addplot[black,dashed,mark=,error bars/.cd, y dir=both, y explicit] table [x=Qubit,y expr=\thisrow{AMT13} / 1000,y error expr=\thisrow{MTStd13} / 1000] {data/tD1_5-0_5-dist.dat};
  \addplot[black,mark=,error bars/.cd, y dir=both, y explicit] table [x=Qubit,y expr=\thisrow{AMT13} / 1000,y error expr=\thisrow{MTStd13} / 1000] {data/tD1_5-0_5-8x8.dat};
  \addplot[brown,mark=,error bars/.cd, y dir=both, y explicit] table [x=Qubit,y expr=\thisrow{AMT13} / 1000,y error expr=\thisrow{MTStd13} / 1000] {data/tD1_5-0_5-k16.dat};
  \addplot[red,mark=,error bars/.cd, y dir=both, y explicit] table [x=Qubit,y expr=\thisrow{AMT13} / 1000,y error expr=\thisrow{MTStd13} / 1000] {data/tD1_5-0_75.dat};
  \addplot[green,mark=,error bars/.cd, y dir=both, y explicit] table [x=Qubit,y expr=\thisrow{AMT13} / 1000,y error expr=\thisrow{MTStd13} / 1000] {data/tD2-0_5.dat};
\end{axis}
\end{tikzpicture}}
\pgfplotsset{height=0.33\textwidth,width=0.3\linewidth,compat=1.9}
\subfloat[]{\label{figures:executionTime13_2}\begin{tikzpicture}
\begin{axis}[
    xlabel={Target Atom Count},ymin=0,xmin=0,xmax=500
    ]
  \addplot[blue,mark=,error bars/.cd, y dir=both, y explicit] table [x=Qubit,y expr=\thisrow{AMT13} / 1000,y error expr=\thisrow{MTStd13} / 1000] {data/timeDemands1_5-0_5.dat};
  \addplot[black,dashed,mark=,error bars/.cd, y dir=both, y explicit] table [x=Qubit,y expr=\thisrow{AMT13} / 1000,y error expr=\thisrow{MTStd13} / 1000] {data/timeDemands1_5-0_5-dist.dat};
  \addplot[black,mark=,error bars/.cd, y dir=both, y explicit] table [x=Qubit,y expr=\thisrow{AMT13} / 1000,y error expr=\thisrow{MTStd13} / 1000] {data/timeDemands1_5-0_5-8x8.dat};
  \addplot[brown,mark=,error bars/.cd, y dir=both, y explicit] table [x=Qubit,y expr=\thisrow{AMT13} / 1000,y error expr=\thisrow{MTStd13} / 1000] {data/timeDemands1_5-0_5-k16.dat};
  \addplot[red,mark=,error bars/.cd, y dir=both, y explicit] table [x=Qubit,y expr=\thisrow{AMT13} / 1000,y error expr=\thisrow{MTStd13} / 1000] {data/timeDemands1_5-0_75.dat};
  \addplot[green,mark=,error bars/.cd, y dir=both, y explicit] table [x=Qubit,y expr=\thisrow{AMT13} / 1000,y error expr=\thisrow{MTStd13} / 1000] {data/timeDemands2-0_5.dat};
\end{axis}
\end{tikzpicture}}
\subfloat[]{\label{figures:executionTime55}\begin{tikzpicture}
\begin{axis}[
    xlabel={Target Atom Count},ymin=0,xmin=0,xmax=3700
    ]
  \addplot[blue,mark=,error bars/.cd, y dir=both, y explicit] table [x=Qubit,y expr=\thisrow{AMT55} / 1000,y error expr=\thisrow{MTStd55} / 1000] {data/default.dat};
  \addplot[black,dashed,mark=,error bars/.cd, y dir=both, y explicit] table [x=Qubit,y expr=\thisrow{AMT55} / 1000,y error expr=\thisrow{MTStd55} / 1000] {data/tD1_5-0_5-dist.dat};
  \addplot[black,mark=,error bars/.cd, y dir=both, y explicit] table [x=Qubit,y expr=\thisrow{AMT55} / 1000,y error expr=\thisrow{MTStd55} / 1000] {data/tD1_5-0_5-8x8.dat};
  \addplot[brown,mark=,error bars/.cd, y dir=both, y explicit] table [x=Qubit,y expr=\thisrow{AMT55} / 1000,y error expr=\thisrow{MTStd55} / 1000] {data/tD1_5-0_5-k16.dat};
  \addplot[red,mark=,error bars/.cd, y dir=both, y explicit] table [x=Qubit,y expr=\thisrow{AMT55} / 1000,y error expr=\thisrow{MTStd55} / 1000] {data/tD1_5-0_75.dat};
  \addplot[green,mark=,error bars/.cd, y dir=both, y explicit] table [x=Qubit,y expr=\thisrow{AMT55} / 1000,y error expr=\thisrow{MTStd55} / 1000] {data/tD2-0_5.dat};
\end{axis}
\end{tikzpicture}}
\caption{Average time to execute the generated rearrangement sequence given a target atom count, a time-demand function, an initial filling ratio, the ratio $r$ between total and target size, and limiting factors to the parallelism. Error bars denote $\pm$ one standard deviation. Shown times are only exemplary and may differ between hardware setups. (a) Complete atom count range; times simulated using $t_m(d)$. (b) Closer view of 0-500 atoms; times simulated using $t_m(d)$. (c) Complete atom count range; times simulated using $t_{m,2}(d)$.\label{figures:executionTime}}
\end{figure*}

In the following, we show the results obtained by using our rearrangement scheme with different parameters. In all figures, we plot results for different numbers of atoms in the target region, different initial filling ratios, and different ratios $r$ between total and target area size. The total array will be of size $\lceil\sqrt{n_t}\cdot r\rceil\bigtimes\lceil\sqrt{n_t}\cdot r\rceil$ and will contain the centered target area with a size of $\sqrt{n_t}\bigtimes\sqrt{n_t}$ sites. All data points are based on 1000 iterations using different randomly generated boolean arrays. In order to improve legibility of the figures, we will use a default configuration and will only note any deviations from it instead of always stating all parameters. That default configuration uses an initial filling ratio $f=0.5$, meaning each site it 50\% likely to be occupied, a total-to-target array ratio $r=1.5$, a limit to the number of frequencies for both vertical and horizontal \ac{AOD} of $n_h=n_v=16$, and a limit to the total number of generated movable traps $k=256$, which effectively means unlimited, as this number is anyway bound by $n_h\cdot n_v=256$. Also, we disallow moving an atom multiple times but allow the movement of empty tweezer onto occupied target positions.

All tests were conducted on a device using an Intel Core i7-1165G7 processor running at 2.8Ghz, 16GB of RAM and Ubuntu 22.04 as the operating system.
\subsection{Characteristics of Move Sequence}
To begin our results, we show the number of moves produces for different parameters in Fig.~\ref{figures:moveCounts}. To compare different fitness functions, we define another one, which uses a faster transport speed and should, therefore, produce moves that prioritize a low move count over total distance.
\begin{equation}
    t_{m,2}(d)=120\mu s+\frac{d}{0.55\frac{\mu m}{\mu s}}
\end{equation}
To generate our data, we mostly used $t_{m,2}(d)$ as a cost function. We can see throughout this analysis that the difference between the two is marginal. In general, we can see that the number of moves is lower for higher initial filling ratios, which makes intuitive sense, as we need to sort less atoms into the target area. Also, larger source arrays don't seem to meaningfully reduce the number of required moves. Limiting both $n_h$ and $n_v$ to 8 increases the number of moves drastically, albeit not by a factor of 2. Altering the default configuration by decreasing the total allowed number of traps $k$ to 16 shows a similar tendency, although not as severe. Most notably, we can see that after an steeper initial rise until around 600-800 target atoms, the number of moves seems to scale approximately linearly with the number of target sites. This is totally expected, since the required number of target sites scales linearly and our algorithm can only parallelize the filling of a given subset of them. The initial steeper rise can easily be explained by the fact that the algorithm cannot fully utilize its parallelization capabilities for smaller array sizes.

\pgfplotsset{height=0.33\textwidth,width=\axisdefaultwidth,compat=1.9}
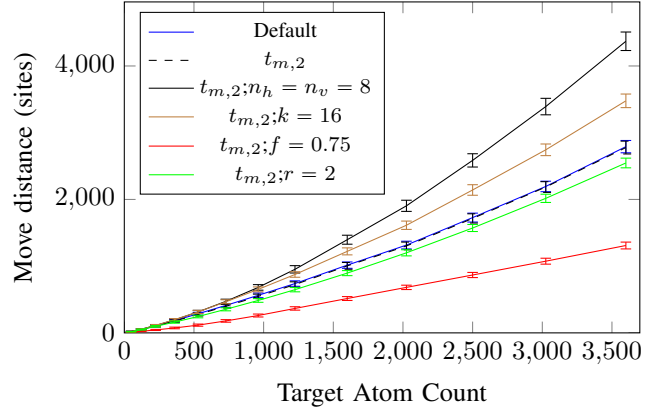
\begin{figure}
\centering
\begin{tikzpicture}
\begin{axis}[
    xlabel={Target Atom Count},
    ylabel=Move distance (sites),
    legend pos=north west,ymin=0,xmin=0,xmax=3700,
    legend entries={Default,$t_{m,2}$,$t_{m,2}\text{;}\textstyle n_h=n_v=8$,$t_{m,2}\text{;}k=16$,$t_{m,2}\text{;}f=0.75$,$t_{m,2}\text{;}r=2$},legend style={font=\footnotesize}
    ]
  \addplot[blue,mark=,error bars/.cd, y dir=both, y explicit] table [x=Qubit,y=AMD,y error=MDStd] {data/tD1_5-0_5-dist.dat};
  \addplot[black,dashed,mark=,error bars/.cd, y dir=both, y explicit] table [x=Qubit,y=AMD,y error=MDStd] {data/default.dat};
  \addplot[black,mark=,error bars/.cd, y dir=both, y explicit] table [x=Qubit,y=AMD,y error=MDStd] {data/tD1_5-0_5-8x8.dat};
  \addplot[brown,mark=,error bars/.cd, y dir=both, y explicit] table [x=Qubit,y=AMD,y error=MDStd] {data/tD1_5-0_5-k16.dat};
  \addplot[red,mark=,error bars/.cd, y dir=both, y explicit] table [x=Qubit,y=AMD,y error=MDStd] {data/tD1_5-0_75.dat};
  \addplot[green,mark=,error bars/.cd, y dir=both, y explicit] table [x=Qubit,y=AMD,y error=MDStd] {data/tD2-0_5.dat};
\end{axis}
\end{tikzpicture}
\caption{Average total distance of moves that our rearrangement algorithm generates given a target atom count, a time-demand function, an initial filling ratio, the ratio $r$ between total and target size, and limiting factors to the parallelism. Error bars denote $\pm$ one standard deviation.\label{figures:totalDist}}
\end{figure}

In Fig.~\ref{figures:totalDist}, which shows the average total move distance, we can draw more of the same conclusions. The only noticeable difference is the different scaling behaviour. Whereas the number of moves increases more steeply for lower atom counts before suddenly following a approximately linear curve, the distance actually increases more than linearly from the beginning.

Fig.~\ref{figures:executionTime} shows the required time to physically execute the generated moves according to our time-demand functions. We do want to note that we do not claim to be able to actually rearrange atoms within the shown time frame. As stated in Section~\ref{section:tdf}, the chosen cost function is used to allow for an easy comparison with alternative algorithms. Using more conservative cost functions, e.g, 800$\mu$s in total for transferring atoms from and to the stationary grid would lead to 6-7 times the execution time. It is also meaningless to compare results of different cost functions among each other, as some simply requires more time. Fig.~\ref{figures:executionTime13} and Fig.~\ref{figures:executionTime13_2} simulate the required execution time using $t_m(d)$, whereas Fig.~\ref{figures:executionTime55} uses $t_{m,2}(d)$. We can see how the results of the more distance-dominated cost function $t_m(d)$ very closely resembles Fig.~\ref{figures:totalDist}. While this does not completely change for Fig.~\ref{figures:executionTime55}, the scaling behaviour clearly starts to exhibit characteristics previously seen for the move count, such as the steeper initial rise.

\subsection{Algorithm Run Time and Success Rate}

\pgfplotsset{height=0.25\textwidth,width=\axisdefaultwidth,compat=1.9}
\begin{figure}
\centering
\begin{tikzpicture}
\begin{axis}[
    xlabel={Target Atom Count},
    ylabel=Success rate,
    legend pos=south east,ymin=0.7,xmin=0,xmax=500,
    legend entries={Algorithm success rate, Chance to contain enough atoms}
    ]
  \addplot[blue,dashed,mark=x] table [x=Qubit,y=SRate] {data/timeDemands1_5-0_5.dat};
  \addplot[red,mark=] table [x=Qubit,y=SuccessRate1.5] {data/theoreticalSuccessRates.dat};
\end{axis}
\end{tikzpicture}
\caption{The success rate of our algorithm along with the chance of the source array containing sufficiently many atoms for $r=1.5$, i.e, the total array has side lengths 1.5 times the side lengths of the target area.\label{figures:successRate}}
\end{figure}
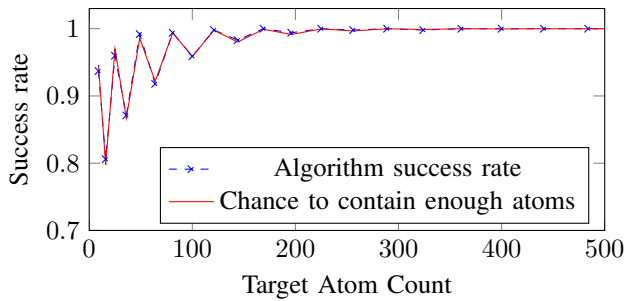

Having investigated the characteristics of the generated rearrangement sequences, we now investigate the success rate of our algorithm. It is designed to always produce a correct sequence of moves if the number of atoms in the total atom array is sufficient. Fig.~\ref{figures:successRate} confirms this by showing that the success rate of the algorithm nearly perfectly coincides with the chance of the array containing enough atoms. The periodic spikes stem from the fact that rounding up the total array size for odd target area side lengths leads to a higher rate of source to target sites.

\pgfplotsset{height=0.33\textwidth,width=\axisdefaultwidth,compat=1.9}
\begin{figure}
\centering
\begin{tikzpicture}
\begin{axis}[
    xlabel={Target Atom Count},
    ylabel=Time (ms),
    legend pos=north west,ymin=0,xmin=0,ymax=100,
    legend entries={Default,$t_{m,2}$,$t_{m,2}\text{;}\textstyle n_h=n_v=8$,$t_{m,2}\text{;}k=16$,$t_{m,2}\text{;}f=0.75$,$t_{m,2}\text{;}r=2$},legend style={font=\footnotesize}
    ]
  \addplot[blue,mark=None] table [x=Qubit,y expr=\thisrow{CTime} / 1000] {data/timeDemands1_5-0_5.dat};
  \addplot[black,dashed,mark=None] table [x=Qubit,y expr=\thisrow{CTime} / 1000] {data/timeDemands1_5-0_5-dist.dat};
  \addplot[black,mark=None] table [x=Qubit,y expr=\thisrow{CTime} / 1000] {data/timeDemands1_5-0_5-8x8.dat};
  \addplot[brown,mark=None] table [x=Qubit,y expr=\thisrow{CTime} / 1000] {data/timeDemands1_5-0_5-k16.dat};
  \addplot[red,mark=None] table [x=Qubit,y expr=\thisrow{CTime} / 1000] {data/timeDemands1_5-0_75.dat};
  \addplot[green,mark=None] table [x=Qubit,y expr=\thisrow{CTime} / 1000] {data/timeDemands2-0_5.dat};
\end{axis}
\end{tikzpicture}
\caption{Average time the algorithm takes to generate the move sequence given a target atom count, a time-demand function, an initial filling ratio, and the ratio $r$ between total and target size\label{figures:runTime}}
\end{figure}
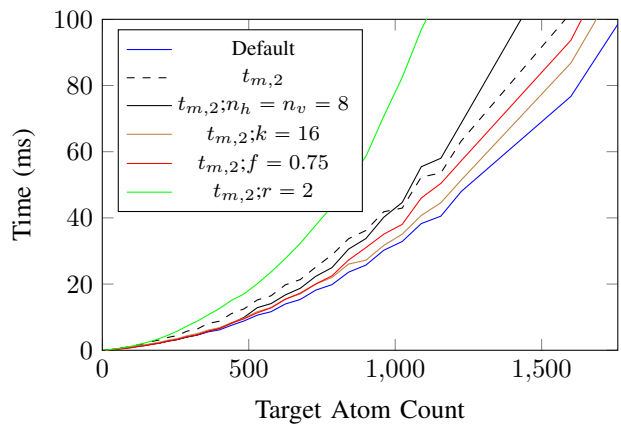

Lastly, there is the run time of our algorithm, which Fig.~\ref{figures:runTime} shows for different target atom numbers and parameters. Surprisingly, we can see that higher initial filling ratios do not decrease run time. Since the algorithm execution only stops once every target site is filled, it effectively executes its main loop as many times as there are moves produced. This explains some differences where we see diverging move counts, such as the increased run time for the less parallel versions. Most notably, the version configured to use $r=2$ takes significantly longer. We attribute this to the fact that the algorithm has a lot more move choices and has to deal with nearly twice as many total atoms, i.e, data.
\section{Related Work}\label{section:relatedWork}
Since we want to discuss in Section~\ref{section:eval} whether our algorithm can offer advantages over existing methods, we first need to evaluate the performance of these known approaches. Atom rearrangement has been a notable topic in the field of cold-atom physics and, more recently, neutral atom quantum computing. Since the time constraint of this process is mostly imposed by the desire to limit a quantum computer's or simulator's compute cycle time, this has become increasingly important with the rise of this technology's popularity. In this section, we discuss some existing methods to rearrange atoms into target configurations.
\subsection{AOD-based Rearrangement}
Since AOD-based rearrangement procedures have usually been the most prevalent method, most research focusses around this approach. There are sequential algorithms, as well as, more recently, parallel ones. 

For atom numbers that we currently deal with, it is usually possible to consider all possible moves in the sequential case. There are several methods of selecting the next move to execute. There are solver-based methods such as solving the \ac{LSAP} \cite{Schymik:1}, some rely on solutions to well-known problem such as the \ac{ASA}, and some use heuristic-based methods such as \ac{HCA} or \ac{HPFA} \cite{Sheng:1,Barredo:1}. 

\begin{figure}
\centering
  \subfloat[]{\includegraphics[width=0.5\linewidth]{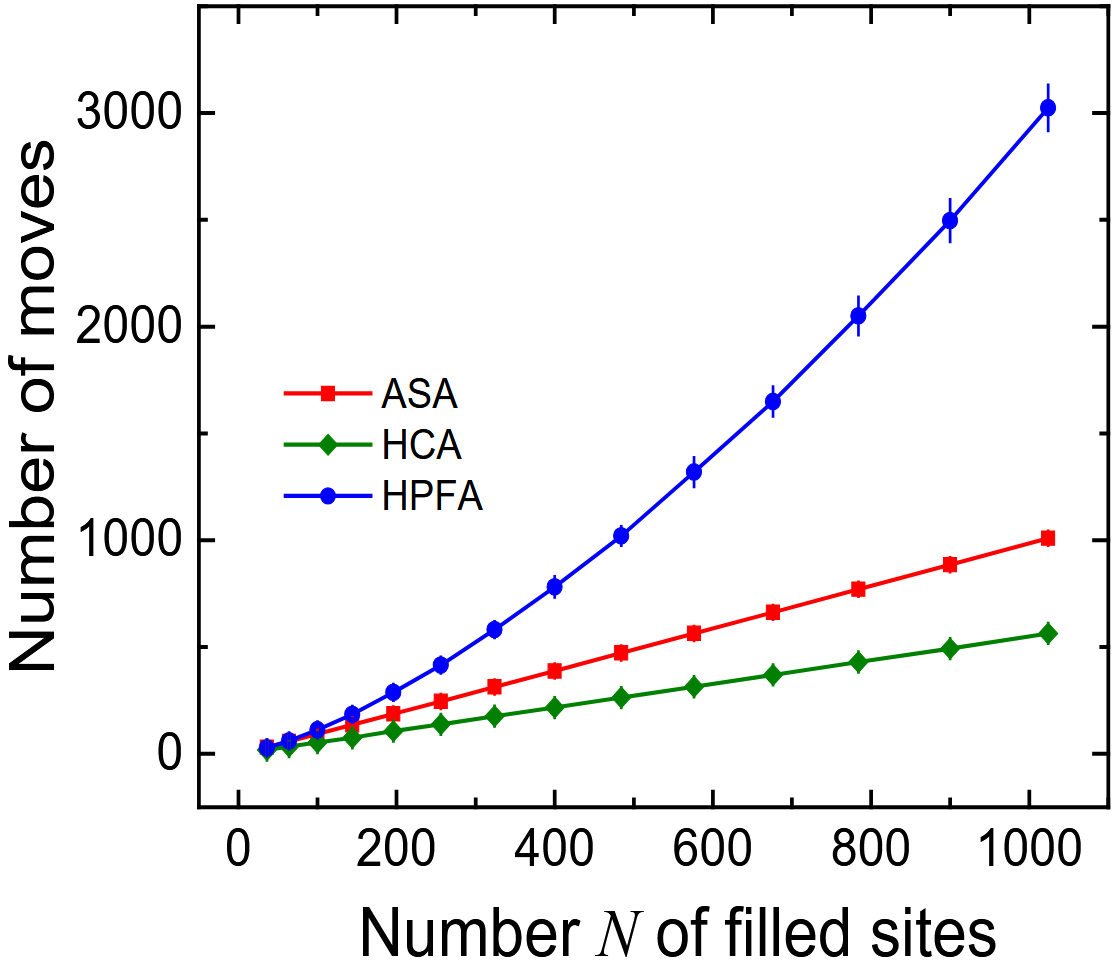}}
  \subfloat[]{\includegraphics[width=0.5\linewidth]{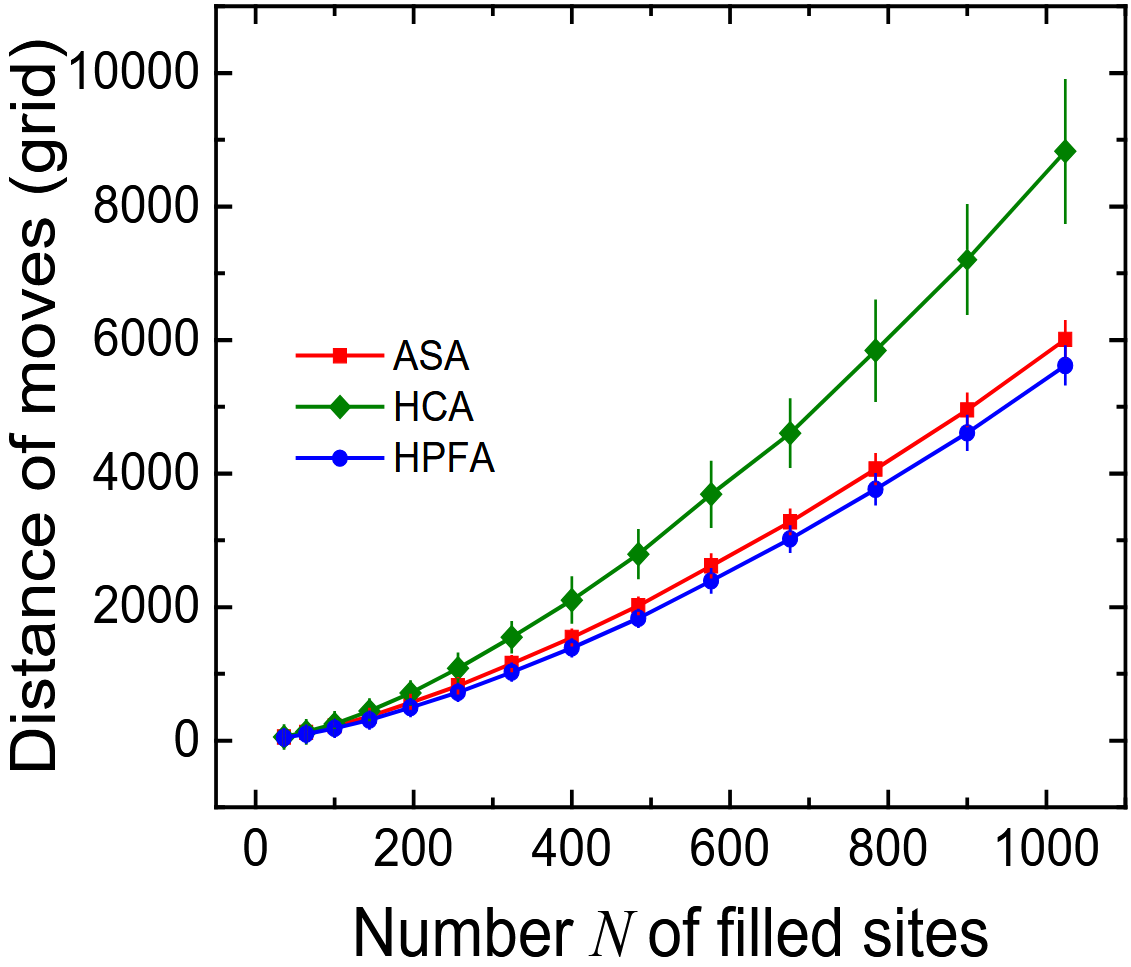}}
\caption{Number of moves of HCA, ASA and HPFA and filling fractions of the defect-free atom array as the number of filled sites. (a) Simulating results of the transfer moves as a function of filled site numbers. (b) Simulating results of the transfer distances as a function of filled site numbers. The error bars are caused by the randomly distributed initial atom arrays with 50\% loading rate. The shape of the target sites are all squares. \\Source: \cite{Sheng:1}}\label{figures:seqComparison}
\end{figure}

Fig.~\ref{figures:seqComparison} shows \citeauthor{Sheng:1}'s results for comparing number of moves and total distance for some of these algorithms \cite{Sheng:1}. Some approaches also include additional imaging steps to check for errors during rearrangement \cite{Mello:1}.

However, as mentioned before, we are capable of feeding several \ac{RF} tones to both the horizontal and vertical \ac{AOD}. This obviously constitutes a huge advantage over sequential approaches, since we can at least execute some, previously sequential, moves simultaneously. Most probably, we can do even better than that and develop an algorithm that is specifically tailored to produce moves that can be executed together. 

After parallel compactification was shown on one-dimensional atom arrays nearly ten years ago \cite{Endres:1}, this idea has matured and evolved into parallel two-dimensional rearrangement schemes for state preparation.

\citeauthor{Ebadi:1} have implemented an algorithm that first shifts atoms horizontally so that each column contains sufficiently many atoms. Subsequently, these are shifted to their target positions in parallel \cite{Ebadi:1}. Their use of comparatively short moves suggests that the focus lies in minimizing total distance as opposed to move count. Regardless, their scheme seems to be well suited for cases of arbitrary target occupancy where the required number of atoms roughly matches the initial number of occupied sites.

\begin{figure}
\centering
  \subfloat[]{\includegraphics[width=0.4\linewidth]{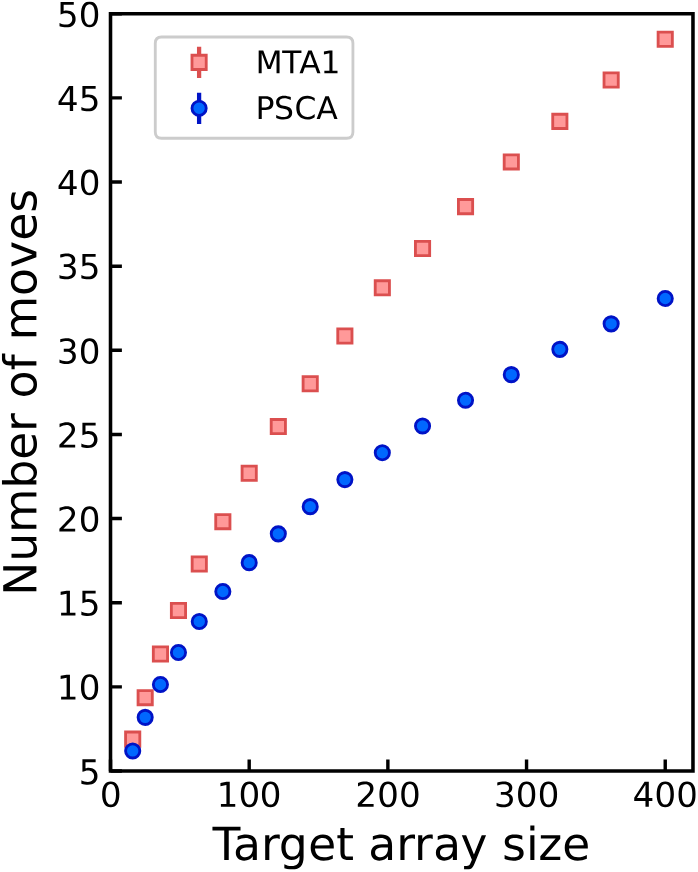}}
  \hspace{1cm}
  \subfloat[]{\includegraphics[width=0.4\linewidth]{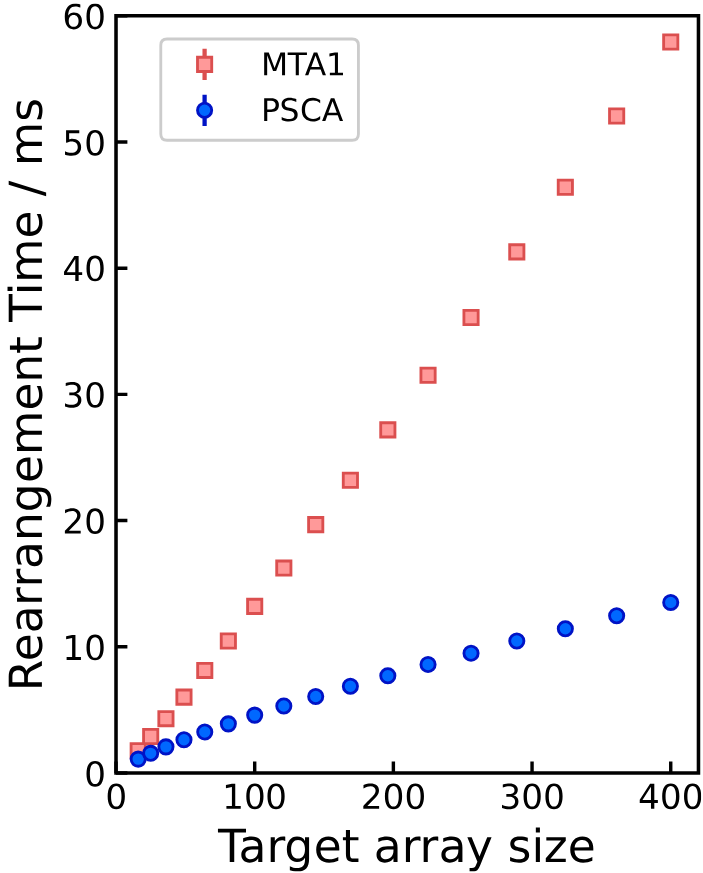}}
\caption{Comparison of the PSCA against MTA1. (a) Simulated scaling of the number of moves for different target array sizes using the PSCA (blue circles) and MTA1 (pink squares). (b) Simulated scaling of the atom rearrangement time for different target array sizes using the PSCA (blue circles) and MTA1 (pink squares). 10,000 iterations of the simulation are run with different random initial loading configurations. The error bars are smaller than the marker size on the plots.\\Source: \cite{Tian:1}}\label{figures:PSCAvsMTA}
\end{figure}

\citeauthor{Tian:1} also propose a parallel movement scheme called \ac{PSCA} and compare it to both \citeauthor{Ebadi:1}'s strategy, which they call MTA1, as well as the single-tweezer \ac{LSAP} approach. Compared to MTA1, they employ a relatively similar strategy of compactifying first in one direction and then the other. Their main advantage seems to be the avoidance of redundant moves \cite{Tian:1}.
\subsection{SLM-based Rearrangement}
With the relatively recent releases of high-refresh-rate \acp{SLM} with resolutions exceeding 1000$\bigtimes$1000, using them for atom rearrangement has become increasingly popular. They generally enable one to fully parallelize the sorting procedure by moving all atoms simultaneously towards their target positions in steps. Usually, a collision free path for each atom is calculated first, which is then interpolated at each step.

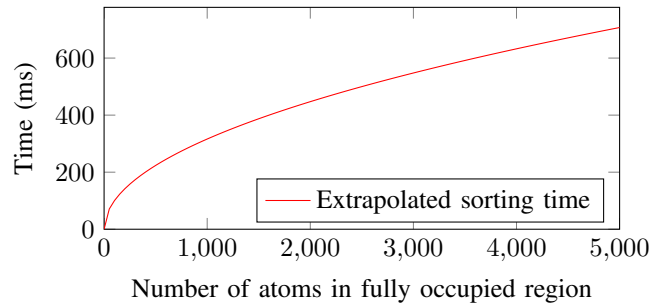
\begin{figure}
\centering
\pgfplotsset{height=0.25\textwidth,width=\axisdefaultwidth,compat=1.9}
\begin{tikzpicture}
\begin{axis}[
    xlabel={Number of atoms in fully occupied region},
    ylabel=Time (ms),
    legend pos=south east,ymin=0,xmin=0,xmax=5000,
    legend entries={Extrapolated sorting time}
    ]
  \addplot[red,samples=100,domain=0:5000] {10*sqrt(x)};
\end{axis}
\end{tikzpicture}
\caption{Extrapolated time spent on rearrangement moves using \citeauthor{Knottnerus:1}'s run time and scaling properties \label{figures:slmRunTime}}
\end{figure}

Fig.~\ref{figures:slmRunTime} shows the approximate time we would expect the sorting to take using \citeauthor{Knottnerus:1}'s time of 40ms for sorting a 6x6 into a fully occupied 4x4 target area and extrapolating it using their claim of scaling with $\mathcal{O}(\sqrt{n_t})$. If we assume a starting grid with 1.5 times the side lengths as the target subgrid, then we can expect the longest move in their setup to be around $\sqrt{2}\cdot\frac{\sqrt{n_t}}{4}$. Since this is exactly the case in the exemplary 4x4 case, we can assume the time of 40ms to be quite representative \cite{Knottnerus:1}.

\citeauthor{Lin:1} also use an SLM-based movement setup. They claim a constant resorting time of under 60ms for 1000 to 10000 qubits with experimental verification of up to 2024 atoms \cite{Lin:1}. This claim is quite surprising, since the distance of their longest individual move should also scale with $\mathcal{O}(\sqrt{n_t})$. Also remarkable is the comparatively low number of $\approx$20 steps in the experimental setup since \citeauthor{Knottnerus:1} already use 13 steps for a target array of 4x4. Considering this low number of steps, perhaps the claim of constant time is based on an observation that for up to 10000 atoms, the execution time of the GPU computations exceeds the total refresh time of the \ac{SLM}.

Like \citeauthor{Lin:1}, \citeauthor{Lee:1} also utilize the Hungarian algorithm to calculate the matching from source to target sites. They also take images throughout the sorting stage to confirm target occupancy or trigger another rearrangement step. This, combined with a comparatively slow \ac{SLM} refresh rate of 200Hz, leads to total sorting times of up to several seconds \cite{Lee:1}. \citeauthor{Kim:1} also use a similar procedure operating at an even lower 60Hz \cite{Kim:1}.

Whether or not \acp{SLM} are actually the best choice for rearrangement right now, it is irrefutable that the recent interest is well-founded. This is especially exacerbated by the expectation that \ac{SLM} technology will probably improve to provide far higher refresh rates in the not-so-distant future \cite{Peng:1, Benea-Chelmus:1}.
\section{Discussion}\label{section:eval}
Having shown the current best alternatives as well as the way our rearrangement sequencing algorithm works and the types of moves it generates along with their characteristics and time demands in terms of both run time and physical execution time, we can now evaluate whether or in which situations our approach can offer a meaningful improvement against established ones.

To begin with, we have to discuss the general limitations of our algorithm and where we see it being useful. It is, in its current form, only capable of producing fully occupied atom arrays. However, an additional presorting step that only uses initially occupied undesired atoms in the final area to fill some unoccupied required sites should alleviate this problem at a relatively small cost, at least in the case where the rate of sites in the target area that we want to be filled greatly exceeds the initial filling ratio. Concerning the hardware setup, it is of great advantage to be capable of moving between rows and columns and to have a large number of frequencies to steer the \acp{AOD}. We do not claim to necessarily offer any advantage should these criteria not be satisfied.

However, in the cases where all advantages of our approach can be used, the resulting rearrangement sequence is far shorter than other investigated \ac{AOD}-based ones with the best of these alternatives showing around $13ms$ of execution time for a target array size of 400 \cite{Tian:1}. Our algorithm can, for this case of $f=0.75$, produce sequences that can be performed within approximately $5.5ms$ using the same time-demand function. Stressing the need for complex moves can produce as little as around 20 moves while the opposite offers the possibility of shifting the balance more towards low distance by altering the configurable cost function.

This cut of over 50\% in the mentioned case of a target array size of 400 should even be a rather minor improvement compared to potential cases where the constant time, which we defined to be $2\cdot60\mu s$, is chosen more conservatively. As stated before, there are also sources putting this time at $2\cdot400\mu s$. In such a case, complicated moves outperform simple shorter ones even more.

We do recognize that the high computation time might be prohibitive for some cases. That being said, we believe that at least up to around 1000 qubits, the time gained through fewer and more complex moves exceeds the additional run time. This is exacerbated by the fact that we do not know the run time for many sources and the ones that we do know are comparable to our approach, e.g, 1.5ms for $f=0.75$, $\sqrt{n_t}=15$ \cite{Tian:1} compared to our $\approx2.8ms$. Also, our algorithm is not fully optimized and runs completely sequential at the moment. This is something that we can fairly easily improve upon in the future to at least run all move-suggesting methods in parallel.

Comparing against \ac{SLM}-based approaches paints a less clear picture. It is quite probable that for atom counts exceeding a few thousand, \ac{AOD}-based approaches will be outscaled by these fully parallelizable ones, especially if future industry breakthroughs can increase the refresh rates of \acp{SLM} further, which is to be expected. However, in the current state of up to around 1000 atoms and with current technology, our approach seems to still be ahead.
\section{Conclusion}
This project was inspired by what we perceived to be untapped potential in the parallelizability of \ac{AOD}-based atom rearrangement procedures. Whereas existing approaches only parallelize in one dimension at a time, we aimed at finding a way of generating highly-parallel moves that utilize multiple \ac{RF} tones for both the vertical and the horizontal \ac{AOD} simultaneously.

To achieve this goal, we developed a heuristic greedy algorithm that, at any stage during rearrangement, investigates different types of moves and select the one that is deemed best. To find the best option, we defined our fitness function as the number of target positions that a move fills divided by the time it takes to execute it. There are, at the moment, four different types of moves that are suggested and that each excel in different conditions such as specific configurations or situations during the rearrangement sequence. Crucially, we also implemented further optimization steps that can improve the best move further by adding additional compatible frequencies. This has actually proven to be quite important for increasing the degree of parallelism that the generated moves exhibit, especially towards the end of the rearrangement procedure.

We have found that our approach produces move sequences that are much faster to execute that any known alternative algorithms. Compared to other \ac{AOD}-based approaches, we can reduce the physical rearrangement time by over 50\% in investigated cases. Comparing against setups using \acp{SLM}, our algorithm, including run time and execution time, can be faster up to around 1000 qubits. For systems larger than that, the \ac{SLM}'s massive parallelization capabilities seem to outperform the \ac{AOD}'s transport speed.

A factor that is currently somewhat limiting our algorithm's usefulness in some regimes is its comparatively long run time. For the default configuration, run time exceeds our simulated execution time for above around 700 qubits and probably prohibits the algorithm's usefulness above around 1500 qubits.

To remedy this limitation, a vital future step is to optimize the algorithm and parallelize the calculations. Also, we want to expand on supported use cases by introducing a presorting steps that clears out undesired atoms in the target area to allow for arbitrary target occupancy matrices. Other potential use cases that we would like to support in the future include other shapes such as triangular or honeycomb configurations and to allow for sorting in lattices where further constraints are imposed.

\section*{Acknowledgments}
We thank the members of the Strontium Rydberg Lab at \ac{MPQ}, \ac{MPQ}'s \ac{MQV} team, and our collaborators at the Fraunhofer Institute for Integrated Circuits for many fruitful discussions on atom rearrangement that allowed us to better understand the underlying physics and hardware constraints.
\bibliography{bibliography.bib}

\end{document}